\documentclass[aps,arl,preprint,superscriptaddress]{revtex4-2}
\pdfoutput=1
\usepackage[utf8]{inputenc} % usado para usuários Linux
\usepackage[T1]{fontenc}
\usepackage[english]{babel} % usado para o português
\usepackage{color}% Controle das cores
\usepackage{graphicx,graphicx}	% Inclusão de gráficos
\usepackage{epsfig,subfig}		% Inclusão de figuras arquivo
\usepackage{dcolumn}% Alinha colunas de tabelas no ponto decimal
\usepackage{bm}% Símbolos matemáticos em negrito
\usepackage{url} 	% Para formatar URLs (endereços da Web)
\usepackage{commath}
\usepackage{color}
\usepackage[usenames,dvipsnames,svgnames,table]{xcolor}
\usepackage[colorlinks=true,linkcolor=black,filecolor=blue,urlcolor=blue,citecolor=blue]{{hyperref}}
\usepackage[labelfont=bf,justification=justified,format=plain]{caption} % 'format=plain' avoids hanging indentation

\usepackage{ulem}

\begin{document}

\title{Experimental verification of the inverse Anomalous spin Hall effect with perpendicular magnetic anisotropy Materials.}

\author{J. E. Abrão}
\email[Corresponding author: ]{elias.abrao@ufpe.br}
\affiliation{\small{Departamento de Física, Universidade Federal de Pernambuco, 50670-901 Recife, PE, Brazil.}}
\author{A. R. Rodrigues}
\affiliation{\small{Departamento de Física, Universidade Federal de Pernambuco, 50670-901 Recife, PE, Brazil.}}

\author{H. F. Ding}
\affiliation{\small{National Laboratory of Solid State Microstructures, Department of Physics, Nanjing University and Collaborative Innovation Center of Advanced Microstructures, Nanjing 210093, People’s Republic of China}}

\author{S. Bedanta}
\affiliation{\small{Laboratory for Nanomagnetism and Magnetic Materials (LNMM), School of Physical Sciences, National Institute of Science Education and Research (NISER), An OCC of Homi Bhabha National Institute (HBNI), Jatni 752050, India.}}

\author{A. Azevedo}
\affiliation{\small{Departamento de Física, Universidade Federal de Pernambuco, 50670-901 Recife, PE, Brazil.}}

\date{December 07, 2023}

\begin{abstract}
In this work, the spin pumping technique was employed to investigate the anomalous inverse spin Hall effect in BIG/NiO/Fe samples where BIG[$(Bi,Tm)_{3}(Fe,Ga)_{5}O_{12}$] exhibits perpendicular magnetic anisotropy. Our results reveal an intriguing phenomenon: when the magnetizations of both ferromagnetic layers align perpendicularly, a distinct spin-to-charge current conversion mechanism occurs. This conversion is intricately linked to the magnetization of the converting layer, spin polarization, and the spin current orientation. 

\end{abstract} 

\maketitle

%\section{Introduction}
Spintronics uses the spin degree of freedom to generate, transport and process information, with spin currents being its central component. A spin current refers to the flow of spin angular momentum. It not only has the ability to transport information\cite{TechnologySpin,Review2}, and exerts torques on magnetic materials\cite{SpinTorque1} and most important, but more importantly, it can also be converted into a charge current.

Over the years, different methods for generating spin currents and enhancing their interaction with charge currents have been discovered. Among these, the most renowned is the spin Hall effect (SHE), where a charge current passing through a metal with large spin orbit coupling (SOC) is partially converted into a transverse spin current. This phenomenon was first proposed by Dyakonov and Perel\cite{Yakonov71} in 1971 and rediscovered by Hirsch\cite{Hirsch99} in 1999.  A reciprocal effect was latter discovered, namely the inverse spin Hall effect (ISHE), where a spin current is converted into a perpendicular charge current allowing the electrical detection of spin current\cite{ISHE-Azevedo,ISHE-Saitoh}. The spin current and charge current densities, generated by SHE and ISHE are described by the following equations,
\begin{equation}\label{SHE-ISHE}
    \begin{split}
    \vec{J^{s}} = \theta_{SH}\frac{\hbar}{2e}\left(\hat{\sigma}\times \Vec{J_{c}} \right),\\\\
    \vec{J^{c}} = \theta_{SH}\frac{2e}{\hbar}\left(\hat{\sigma}\times \Vec{J_{s}} \right),
    \end{split}
\end{equation}
where $\hat{\sigma}$ is the spin polarization and $\theta_{SH}$ is the spin Hall angle, a parameter which determines the efficiency of the interplay between spin and charge current. In Eqs.(\ref{SHE-ISHE}), $\Vec{J}_{s}$ is written as a vector quantity, but as it represents an angular momentum flow, it is rank two tensor. As written, the equations \ref{SHE-ISHE}, which are appropriately employed for understanding experimental data, have been subjected to extensive testing with semiconductors\cite{SHE-SemiCon2,SHE-SemiCon} as well as paramagnetic metals, such as Platinum, Tantalum, Palladium and Tungsten \cite{Joaquim2,Joaquim5,Joaquim1}.

The study of spin-to-charge conversion in ferromagnetic materials (FM) came soon after the seminal investigations of SHE. Initially, it was observed that FM materials not only present SHE and ISHE, but these effects can also be significantly dependent on the specific FM, even compared to a standard material like Pt\cite{ISHE-FM}. Furthermore, another study revealed that the ISHE, in a FM/FM structure remains independent on the magnetization direction of the FM layer responsible for conversion\cite{ISHE-FM-2}. Recently, a theoretical study\cite{AISHE-Teorico} predicted an anomalous spin Hall effect by expanding the spin Hall angle to a rank 3 tensor and considering that the converting material has an order parameter, such as the magnetization $\hat{M}$. In doing that, the spin current and the charge current generated via direct and inverse spin Hall effect gain a new term which is dependent on the order parameter. The spin current and charge current due to the SHE and ISHE are then described by:

\begin{equation}\label{Eq.SHE-ISHE-Anom}
    \begin{split}
        \vec{J^{s}} = \theta_{SH}\frac{\hbar}{2e}\left(\hat{\sigma}\times \Vec{J_{c}} \right) + \theta_{SH}^{A}\frac{\hbar}{2e}\left(\hat{\sigma}\times\left(\hat{M}\times \Vec{J_{c}} \right)\right),\\\\
        \vec{J^{c}} = \theta_{SH}\frac{2e}{\hbar}\left(\hat{\sigma}\times \Vec{J_{s}} \right) + \theta_{SH}^{A}\frac{2e}{\hbar}\left(\hat{\sigma}\times\left(\hat{M}\times \Vec{J_{s}} \right)\right).
    \end{split}
\end{equation}
Where the first terms represent the standard SHE/ISHE effects, while the second terms are the anomalous effects due to the order parameter $\hat{M}$ of the detecting layer. Here, $\theta_{SH}^{A}$ represents the anomalous spin Hall angle. From Eq.(\ref{Eq.SHE-ISHE-Anom}), for a FM/FM system, an extra charge current is converted when the magnetization of the converting FM layer is perpendicular to the direction and polarization of the spin current. Experimental results based on this theory have been demonstrated\cite{AISHE-Exp} using a $YIG/[Co/Pd]_{5}$ heterostructure. In this configuration the $\vec{M}_{YIG}$ is kept fixed in-plane, while the magnetization of $[Co/Pd]$ system is allowed to rotate. Interestingly, it was observed that when the [Co/Pd] layer presents out of plane magnetization, an additional charge current is detected. Since $\vec{M}_{YIG}$ is kept in-plane, components of the anomalous SHE and ISHE, which require an injector with perpendicular magnetic anisotropy (PMA), could not be detected. So far, experiments related to the anomalous spin Hall effects with perpendicularly polarized spin currents are lacking.

In this work, we explore the anomalous inverse spin Hall effect (AISHE) via the spin pumping technique using a ferromagnetic insulator that exhibits perpendicular magnetic anisotropy as injector. Our samples consists of Bismuth-Doped Thulium Iron Garnet [$(Bi,Tm)_{3}(Fe,Ga)_{5}O_{12}$] films grown onto Gadolinium Gallium Garnet substrates (GGG [111]), by the liquid phase epitaxy technique. High purity oxides of $Bi_{2}O_{3}$, $Tm_{2}O_{3}$, $Ga_{2}O_{3}$, $Fe_{2}O_{3}$ and $PbO$ were mixed and melted in a Pt crucible at approximately $1000^{\circ}$C, then cooled down to the growth temperature of around $900^{\circ}$C. Subsequently, we fabricated (BIG)/NiO/Fe heterostructures using sputtering, in which the thin NiO layer allows the flow of a pure spin current and magnetically decouples both ferromagnets. The spin polarization $\hat{\sigma}$ of the pure spin current is set by the direction of the BIG magnetization $\vec{M}_{BIG}$ while the magnetization of the iron layer $\vec{M}_{Fe}$ can be made perpendicular to $\vec{M}_{BIG}$ by rotating the sample. We show that AISHE exists in materials with perpendicular anisotropy, it depends not only on the magnetization of the converting layer but also on the spin current pumped into it, validating Eq.(\ref{Eq.SHE-ISHE-Anom}) to materials with PMA. 

%\section*{Sample Fabrication and Experimental Set-up}

We cut $3\times 2$ $mm^2$ samples from a BIG wafer using a low-speed diamond saw. These samples were subsequently cleaned in an ultrasonic bath containing acetone and isopropyl alcohol before being transferred to the sputtering chamber. The base pressure was maintained at $1.5\times 10^{-7}$ torr or lower throughout the process. All materials where deposited in an Ar atmosphere with a working pressure of 3.0 mtorr. Initially, a 2nm $NiO$ layer was deposited via rf sputtering onto the BIG, followed by the deposition of a 20nm layer of $Fe$ through DC sputtering. The $NiO$ layer was used to decouple the magnetization while still allowing most of the spin current to flow through.

The sample was affixed to the tip of a PVC rod and carefully positioned within an X-band microwave cavity, where the radiofrequency (rf) magnetic field is maximum and the rf electric field is minimum. The rectangular cavity, operating in TE102 mode, was placed between the poles of an electromagnet. This electromagnet generated a DC magnetic field, $H_{0}$ to induce the Ferromagnetic Resonance (FMR) condition, while the rf field was set perpendicular to $H_{0}$. Field scans of the derivative of the absorption power (dP/dH), were obtained via lock-in amplifier by modulating the DC field with a small sinusoidal field at 100 kHz while keeping the rf frequency fixed at 9.5 GHz. To investigate the spin-to-charge conversion, we employed the spin pumping technique. This technique harnesses the dynamics of ferromagnetic material to inject a pure spin current into the adjacent layer when the sample is in the condition of ferromagnetic resonance. The spin pumping signal was detected by attaching two electrodes at the edges of the sample with silver paint, and these electrodes were connected directly to a nanovoltmeter.

%\section*{Results and Discussion}

\begin{figure}[ht]
    \centering
    \includegraphics[width=1.1\linewidth]{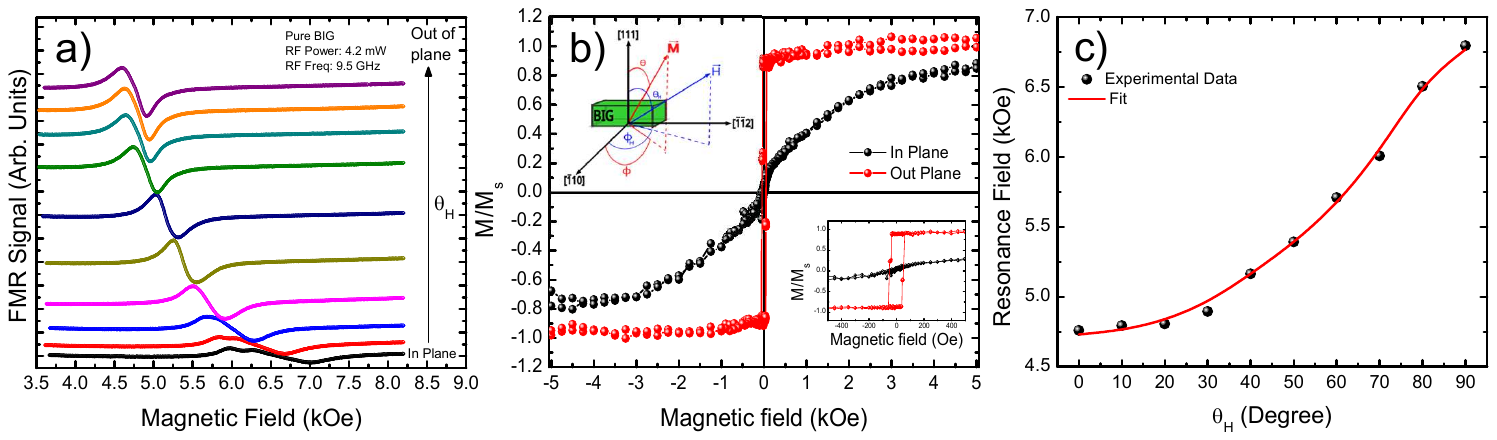}
    \caption{In (a) Field scan of the FMR absorption derivative for different out-of-plane angles for pure BIG. (b) In-plane (black) and out-of-plane (red) magnetization curves of the pure BIG, confirming that the easy axis is perpendicular to the sample plane. (c) Dependence of the FMR field values as a function of $\theta_H$ (blue symbols), the solid line is the best fit obtained for the FMR condition.}
    \label{fig:fig1}
\end{figure}

Fig.(\ref{fig:fig1}) shows the angular dependence of the FMR signal for a pure BIG sample, as the sample is rotated, the direction of the external magnetic field changes from in plane to out-of-plane. The FMR scans show that the resonance field decreases as the direction of the applied field varies from to out-of-plane, this indicates that the BIG indeed has PMA making it a prime candidate to explore the anomalous spin-Hall effect in magnetic materials with in-plane anisotropy. In the out-of-plane configuration, the FMR spectra presents a single well define peak with resonance field $H_{R} = 4.79$ kOe and linewidth $\Delta H = 254$ Oe. Fig.(\ref{fig:fig1}b) shows magnetization curves, obtained by vibrating sample magnetometer, with the magnetic field applied parallel (black symbols) and perpendicular (red symbols) to the sample plane. In Fig. (\ref{fig:fig1}c), the FMR field dependence is depicted as function of $\theta_H$ (black symbols), while the solid line illustrates the best fit based on the FMR condition. This fit takes into account the most relevant terms of the magnetic free energy, for further details see Ref\cite{APL-TIG}.From the numerical fit we extracted the physical parameters for the BIG film, shown in Tab.(\ref{tab:Table1}).

\begin{table}[ht]
    \centering
    \caption{Physical parameters extracted from the best fit shown in Fig. 1(c), where $4\pi M_{eff}$ is the effective magnetization, $H_{1C}$ is the first order cubic anisotropy field, and $H_{U2}$ and $H_{U4}$ are the first and second order uniaxial anisotropy fields, respectively. $H_{U2}$ is the out-of-plane uniaxial anisotropy field, also named $H_{\perp}$.}
    \begin{tabular}{||c|c||}
         \hline
         $4\pi M_{eff} $& $-1670$ G  \\
         \hline
         $H_{1C} = (2K_{1}/M_{s})$ & $-1738$ Oe\\
         \hline
         $H_{U2} = (4\pi M_{eff} - 4\pi M_{s})$& $-2940$ Oe \\
         \hline
         $H_{U4} = 4K_{4}^{\perp}/M_{s}$ &  $-2917$ Oe \\
         \hline
    \end{tabular}
    \label{tab:Table1}
\end{table}

\begin{figure}[ht]
    \centering
    \includegraphics[width=1.0\linewidth]{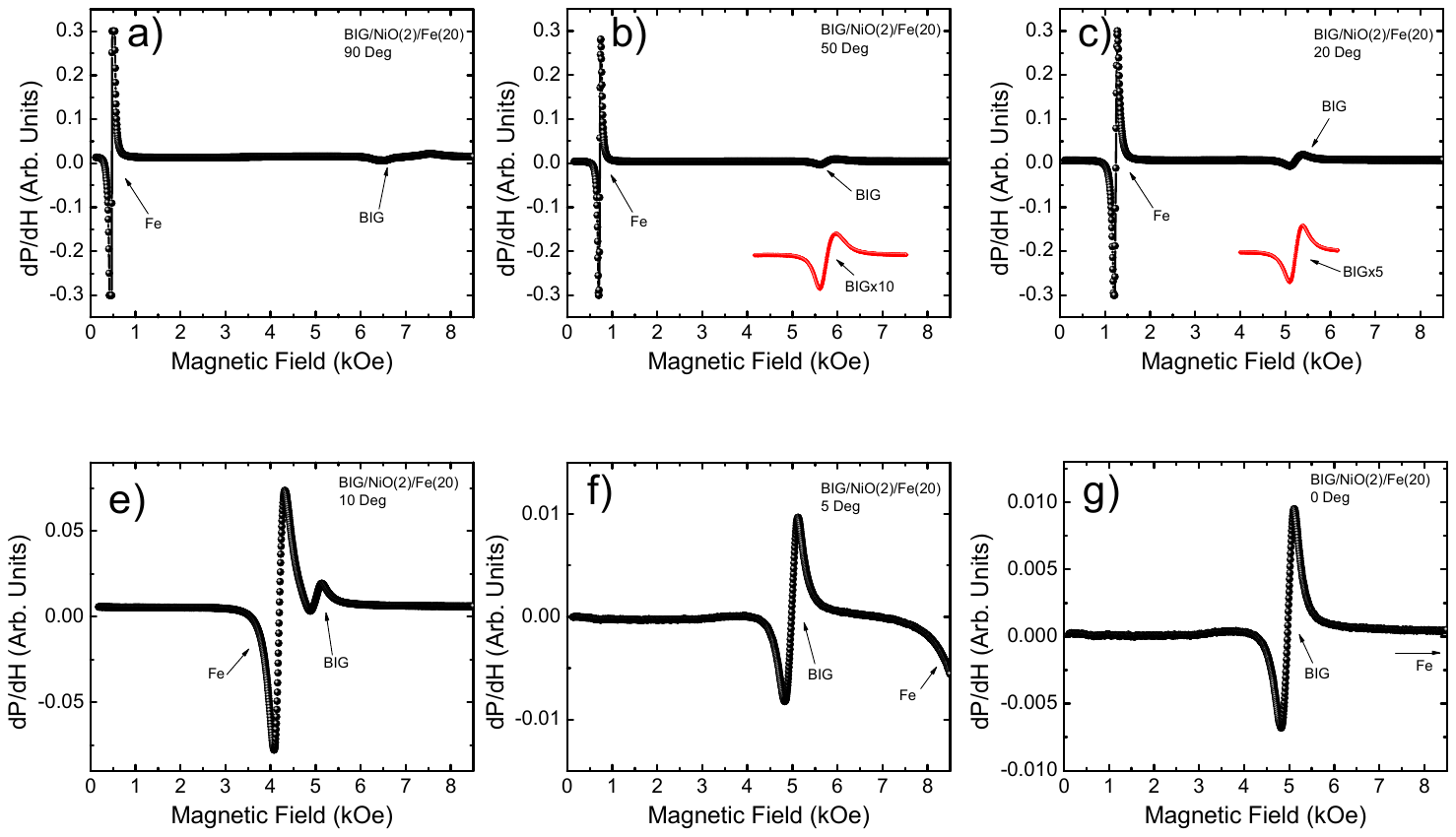}
    \caption{Angular dependence of the FMR Signal for BIG/NiO(2)/Fe(20), in (a) the sample is in the in-plane field configuration i.e $\theta_{H} = 90$º while in (f) is presented the out-of-plane field configuration i.e $\theta_{H} = 0$º. Figures (b) to (e) show intermediary angles. In all cases the RF power used to promote the resonance was $4.2$ mW.}
    \label{fig:fig2}
\end{figure}

After confirming that BIG indeed possesses PMA, the next step was to verify whether the NiO layer effectively decoupled both FM materials. Fig.(\ref{fig:fig2}) shows the angular dependence of the FMR signal for a BIG/NiO(2nm)/Fe(20nm) heterostructure. When the sample is rotated from in-plane to out-of-plane, the resonance field corresponding to the Fe increases while the the signal related to the BIG decreases. Notably, at around $\theta_{H} = 10º$ (Fig. \ref{fig:fig2}f) both signals overlap.However, at 90 degrees the signal related Fe become invisible within the scan range due to the high demagnetization field of the iron film, around $22$ kOe. Consequently, by rotating the sample, it is possible to select the desired field range, enabling the alignment of the desired magnetization with the external magnetic field.

\begin{figure}[ht]
    \centering
    \includegraphics[width=1.0\linewidth]{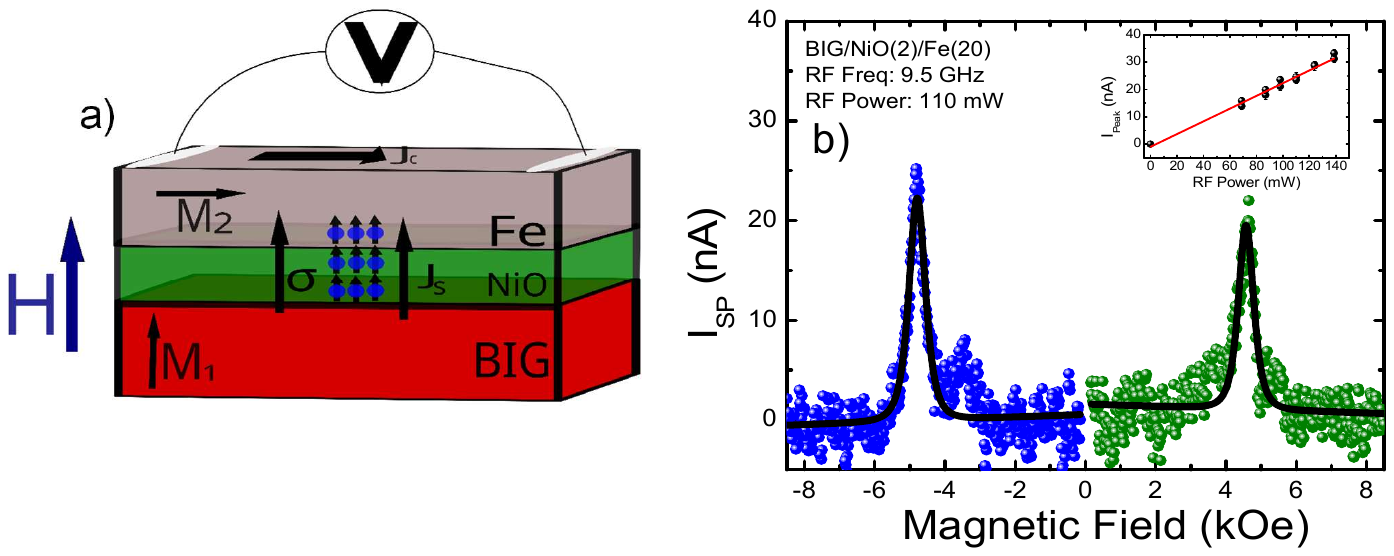}
    \caption{In (a) it is shown an schematic representation of the out-of-plane spin pumping configuration, in this situation the direction of the spin current and its polarization are along the magnetic field. In (b) it is presented the spin pumping signal for $BIG/NiO(2)/Fe(20)$ measured as a function of the out-of-plane field. The inset shows the dependency of the peak current detected via spin pumping with the rf power used to excite the ferromagnetic resonance. The black solid lines are numerical fits to the data using a Lorentzian function.}
    \label{fig:fig3}
\end{figure}

Spin pumping measurements were conducted with the sample fixed in the out-of-plane configuration. In this setup, both the direction of the spin current  ($\vec{J_{s}}$) and its polarization $\hat{\sigma}$ align with the external magnetic field, while $\vec{M_{Fe}}$ remains in-plane. A schematic representation of the system is depicted in Fig.(\ref{fig:fig3}a).

Figs.(\ref{fig:fig3}b) shows the spin pumping measurements conducted on BIG/NiO(2)/Fe(20) under positive and negative perpendicular magnetic field scans. The results reveal that the two distinct peaks exhibit identical polarization, being positively oriented. Due to the fact that $\hat{\sigma}\parallel \vec{J_{s}}$, , the SP signal cannot be explained by Eq.(\ref{SHE-ISHE}) meaning that, the observed signal is not the usual ISHE, however it is consistent with the AISHE, since $\vec{J}_{s}\perp \vec{M}_{Fe}$. By reversing the direction of the perpendicular magnetic field the spin current maintains an upward direction, but the spin polarization changes following the field. Thus, one would expect a reversal of the SP peaks, but this was not observed. Our result indicates that by reversing the field direction, the in-plane magnetization of the iron film also reverses, which agrees with Eq.(\ref{Eq.SHE-ISHE-Anom}). To elucidate the role of the spin current in the processes, measurements were made varying the RF power. The results presented in the inset of Fig.(\ref{fig:fig3}c) show that the system responds linearly to the spin current. 

\begin{figure}[h]
    \centering
    \includegraphics[width=1.0\linewidth]{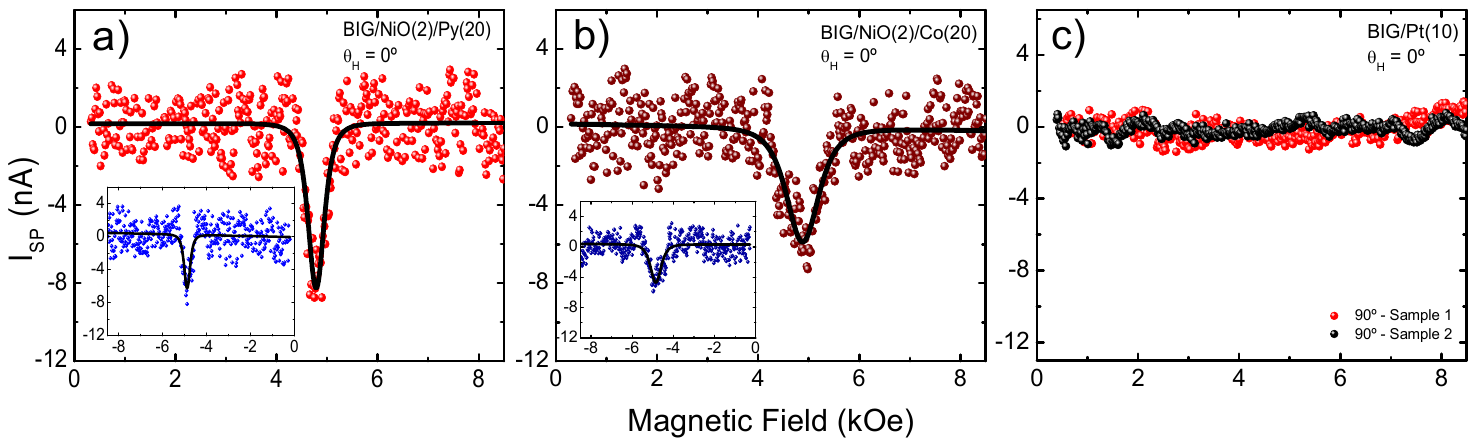}
    \caption{Spin pumping measurements of the anomalous inverse Hall effect, using the same configuration used to obtain the data of Fig. 3(b), for:(a) BIG/NiO(2)/Py(20), (b) BIG/NiO(2)/Co(20) and (c) BIG/Pt(10). The insert in (a) and (b) show the measurement with negative magnetic field, the black solid lines are numerical fits to the data using a Lorentzian function. All measurements where made with a RF Power of 110 mW.}
    \label{fig:fig4}
\end{figure}

To ensure the effect in question is due to the magnetization of the converter layer and not to a particular case of material selection, samples of BIG/NiO(2)/Py(20), BIG/NiO(2)/Co(20) and BIG/Pt(10), where Permalloy (Py) = $Ni_{81}Fe_{19}$. The new samples were placed in the out-of-plane field configuration and spin pumping curves were obtained Fig.(\ref{fig:fig4}) shows the spin pumping signal obtained from  the three new heterostructures. As expected, a SP signal was observed for Py and Co, which were attribute to the AISHE. Interestingly, the signals for Py and Co exhibit an opposite trend to that of Fe, indicating that $\theta_{SH}^{A}$ of $Co$ and $Py$ are opposite to the one of $Fe$. It is worth noting that we refrain from assigning a positive or negative designation to $\theta_{SH}^{A}$ due to an absence of a universally accepted standard, similar to the case of Pt in ISHE. 

To rigorously test our accomplishments, we investigate the SP signal in two different BIG/Pt(10) bilayers. The results shown in Fig.(\ref{fig:fig4}c) shows no observed SP signal, strongly endorsing our hypothesis that the anomalous inverse spin Hall effect obeys the Eq.(\ref{Eq.SHE-ISHE-Anom}), this result is not only consistent with previous works\cite{APL-TIG} but also confirm that the previous observed signal depends on the magnetization of the converting layer. Furthermore, these results affirm that the previous observed signal depends on the magnetization of the converter layer. It is important to mention that, when taking in account symmetries of the system, a third term can been proposed for Eq.(\ref{Eq.SHE-ISHE-Anom})\cite{FMR-MaterialsPerspectives}, however we were unable to detect it due to the geometry used in our system, to detect this extra signal a ferromagnetic detector layer with a magnetization fixed along an arbitrary polar angle between $0^{\circ}$ and $90^{\circ}$ would be needed.

Our findings lead to the conclusion that in a heterostructure consisting of two uncoupled ferromagnetic layers ($FM_{1}/FM_{2}$), the occurrence of anomalous Hall effect is feasible when both magnetizations are perpendicular. Specifically, in cases where the $FM_{1}$ injector possesses perpendicular magnetic anisotropy, an Anomalous Inverse Spin Hall Effect (AISHE) manifests when utilizing $FM_{2}$ materials with in-plane anisotropy for spin current conversion. Certainly, this discovery broadens the spectrum of possibilities for devices seeking to explore the interaction between spin and charge currents.

\section*{Acknowledgements}

This research was supported by Conselho Nacional de Desenvolvimento Científico e Tecnológico (CNPq), Coordenação de Aperfeiçoamento de Pessoal de Nível Superior (CAPES), Fundação de Amparo à Ciência e Tecnologia do Estado de Pernambuco (FACEPE), Universidade Federal de Pernambuco, Multiuser Laboratory Facilities of DF-UFPE, INCT of Spintronics and Advanced Magnetic Nanostructures (INCT-SpinNanoMag), No. CNPq 406836/2022-1. 

\section*{Data Availability}

The data that support the findings of this study
are available from the corresponding author
upon reasonable request.

\bibliography{Refs}% Produces the bibliography via BibTeX.

\end{document}